# Generation of synthetic CT images from optical scanning for superficial mold brachytherapy

Scott Crowe[1,2,3,4][0000-0001-7028-6452], Emily Simpson-Page[1][0000-0002-5262-6465], Jenna Luscombe[1], Rachael Wilks[1,2,3][0000-0002-2110-9676], and Tanya Kairn[1,2,3,4][0000-0002-2136-6138]

[1] Royal Brisbane & Women's Hospital, Herston QLD 4029, Australia
[2] Herston Biofabrication Institute, Herston QLD 4029, Australia
[3] University of Queensland, St. Lucia QLD 4072, Australia
[4] Queensland University of Technology, Brisbane QLD 4001, Australia
sb.crowe@gmail.com

**Abstract.** Optical 3D scanning systems allow the acquisition of accurate models of patient anatomy, suitable for use in the design of simple 3D-printable patient-matched medical devices with 3D modelling software. This study developed and demonstrated the use of superficial brachytherapy surface mold design workflow that utilizes data from optical 3D surface scanning and enables a commercial brachytherapy treatment planning system to be used for catheter positioning and dose optimization steps. Synthetic CT images were generated from 14 optically scanned anatomical models of human participants. Models and skin textures acquired from the optical scans were imported into Autodesk Meshmixer, where the treatment area was delineated, and treatment and device volumes produced. 3D Slicer was used to convert the body, treatment and device volumes to DICOM CT and RTSTRUCT data. The synthetic CT data and contoured volumes were imported into Varian Eclipse, where catheters were designed, and dwell positions and times optimised for dose coverage of the treatment volume. The lack of internal anatomy did not compromise dose calculations, due to clinical use of a TG43 based algorithm. Once 3D printed, molds can be imaged in-situ during CT simulation, and reconstructed, for clinical dose calculation and plan approval.

## 1 Introduction

Superficial brachytherapy is an effective treatment modality for skin cancer and other superficial diseases, with advantages including convenient delivery over fewer fractions than external beam radiotherapy and a rapid dose fall-off sparing underlying and surrounding tissue [1]. Surface molds allow the reproducible placement of brachytherapy sources on flat and irregular anatomical topography [1] and are particularly useful in the treatment of smaller lesions and on complex sites such the face [2] and hand, where the fit (and therefore dosimetry) of traditional surface flaps may be compromised.

In recent years, surface molds have increasingly been produced using 3D printing or additive manufacturing technology. The ability to produce these molds without alginate or plaster casting, using medical images of the patient, reduces patient discomfort and preparation time, and can result in an improved device fit and more homogeneous skin dose [3-4].

As summarized in a 2021 review published by Bellis et al. [4], a majority of publications that have reported on the use of 3D printed surface molds have utilized computed tomography (CT) images for the patient in the device design. However, the use of 3D surface scanning techniques is increasingly being adopted [4-11]. These solutions characterize the skin surface and visible anatomical topography using a non-ionizing optical signal and are typically inexpensive and efficient to use [4-5].

The two most common techniques used for acquiring 3D surface images for the design of brachytherapy molds are structured light scanning [5-6], where the surface topography is characterized by how an incident pattern of light is deformed, and photogrammetry [7-9], where common points or landmarks visible in multiple photographic images are correlated and used to produce a cloud of points in 3D space.

Subsequent to the acquisition of a 3D models of the patient, the devices are then designed with hard surface or parametric modelling tools, such as Blender (Blender Foundation, Netherlands), Meshmixer (Autodesk, USA) or SolidWorks (Dassault Systemes, France) [4]. This generally involves extrusions (or expansions) from existing surfaces to create uniform thickness volumes for the mold.

One disadvantage of a workflow involving only optically scanned data and 3D modelling software is that it does not facilitate any dosimetric optimization of the device design. For example, the positioning of catheter channels within the mold has an impact on the potential quality of dose to the lesion and sparing of normal tissue, which can be more easily assessed within a treatment planning system using isodose distributions and dose volume histograms. Such calculations generally require a CT dataset and segmented treatment volumes, as most treatment planning systems do not support common 3D modelling file formats. For TG43-based dose calculation algorithms [12], this CT data does not need to have include accurate Hounsfield unit assignments.

The generation of simple water-equivalent synthetic CT datasets from surface scans has been described by Skinner et al. [13] and Crowe et al. [6]. Optical scanning systems allow the capture of skin textures, from which marked treatment areas may be segmented and converted to structures within the synthetic CT dataset coordinate system [6].

The objective of this study was to utilize these established approaches in the development and demonstration of a superficial brachytherapy surface mold design workflow using a brachytherapy treatment planning system.

# 2 Method

## 2.1 Optical scanning

The optical scans featured in this study were acquired using an Artec Leo wireless handheld 3D scanner (Artec 3D, Luxembourg), a metrology-grade scanner with features that make it suitable for clinical use in radiotherapy [6,14]. The commissioning and use of the Artec Leo scanner for the scanning of a cohort of volunteers placed in radiotherapy treatment positions has been described by Crowe et al. [6]. The present work utilizes scans acquired of 5 volunteers with approval of the institutional Human Research Ethics Committee.

Prior to scanning, the 5 volunteers had a total of 14 mock treatment areas marked with red marker pen by radiation therapy and medical physics staff. The marked treatment areas were located on the forehead (n=2), nose (n=2), lower eyelid (n=2), upper arm (n=1), lower arm (n=4), finger (n=1), abdomen (n=1) and lower leg (n=1). Depending on the anatomy being scanned, the volunteers were asked to lay supine or sit upright during image acquisition.

Scans were reconstructed using Artec Studio (v15, Artec 3D, Luxembourg) using default reconstruction parameters. The models were cropped to only include the region of interest and surrounding anatomy, achieving a suitable “body” volume for subsequent treatment planning. The 3D models were exported using the Wavefront OBJ file format, in addition to companion material template library files (MTL) and JPEG images, defining the surface texture mapping. These surface textures would allow delineation of the marked mock treatment area on the 3D model in 3D modelling software. Models were also exported using the 3D Systems STL file format.

### 2.2 Data conversion

To facilitate the use of optically scanned 3D model data within the treatment planning system, the body and mock treatment volumes needed to be converted to the DICOM format, using an approach summarized in earlier work [6,13].

The treatment volumes were produced from the surface texture mapped OBJ model, using Autodesk Meshmixer (v3.5.474, Autodesk Inc., United States). This was achieved by delineation of the visible surface treatment area marking using the lasso selection tool, smoothing of the selection boundary to reduce any irregularity in the delineation due to underlying 3D mesh, separation of this treatment area from the surrounding body surface, and finally extrusion of the separated treatment area to a volume using a negative offset corresponding to the dose prescription depth, defined as 5 mm, per GEC-ESTRO ACROP recommendations for skin brachytherapy [1]. The treatment volume was then exported from Meshmixer using the 3D Systems STL file format.

A mock surface mold applicator was also produced using the same approach, but with an additional margin added to the marked treatment area during delineation to accommodate the device itself, and with a positive extrusion offset of 10 mm, to produce a 10 mm thick surface mold. This was also exported using the 3D Systems STL file format.

The STL files corresponding to the body volume, treatment volume and surface mold were imported into the 3D Slicer software (v4.13.0, The Slicer Community) as segmentations. To export these segmentations using the DICOM RTSTRUCT format (widely supported by brachytherapy treatment planning systems), these segmentations needed to be associated with a 3D array of voxels. This was achieved by converting the body segmentation data to a binary labelmap volume, in which voxels enclosed by the body segmentation were assigned a value of 1 and voxels outside the body segmentation were assigned a value of 0. The resolution of this dataset could be modified by changes to the oversampling factor associated with this conversion. For this study a uniform voxel resolution of $0.5 \times 0.5 \times 0.5$ mm$^3$ was used. Once converted, the three segmentations were exported using the DICOM RTSTRUCT format, and the binary labelmap volume exported using the DICOM CT format.

A Hounsfield scale was achieved in these exported DICOM CT files by adjusting the rescale intercept and slope DICOM attributes using pydicom [15], to -1000 and 1000 respectively, resulting in pixel values of -1000 HU for air and 0 for water (instead of 0 and 1).

### 2.3 Treatment planning

The DICOM CT and DICOM RTSTRUCT files were verified to have been created correctly by importing within the MIM Maestro software (MIM Software Inc., USA), used within the department for segmentation prior to treatment planning. DICOM files for a face, upper arm, lower arm, and abdominal mock treatment were subsequently imported into the Varian Eclipse treatment planning system (v13.6, Varian Medical Systems, USA) for planning.

Catheter channels were defined within the existing mold volume manually by the user, at a standoff distance of approximately 5 mm from the skin surface. For molds containing more than one catheter channel, catheters were separated by approximately 5 mm. Image rotation and catheter duplication tools within the Eclipse treatment planning system were used to improve the efficiency of this design process. Dwell positions were automatically generated along the defined catheters, within the mold volume.

The dwell times were optimized to achieve a prescription dose at a depth of 5 mm, corresponding to the depth of the defined treatment volume. Both graphical optimization and TG43-based volume optimization were used to achieve this coverage. In the case of volume-based optimization, dose sparing objectives were defined for both the body volume and a normal tissue optimization volume beneath the existing treatment volume (to push the optimizer to achieve a suitable dose fall-off).

The resultant treatment plan and dose distributions were exported as files containing RTPLAN and RTDOSE objects. Blender was used to produce catheter channel volumes using the dwell position geometries contained in the exported RTPLAN files. Within Blender, each catheter path was defined using a non-uniform rational basis spline (NURBS) open 3D curve, with control point coordinates set to recorded dwell positions. The catheter channel volumes were created by deforming and multiplying a cylindrical mesh (using the curve and array modifiers). These volumes were subtracted from the mold device using boolean modifiers, resulting in designs ready to be 3D printed.

# 3 Results

The average time per scan was 3±1 min and there were no errors in image acquisition (e.g. errors that might require patient re-imaging, in a clinical setting). Despite a variation in skin tones, the red marker pen could be resolved and used for contouring without any optimization of windowing or reconstruction setting.

The conversion between an example lower eyelid reconstructed 3D model and DICOM CT and RTSTRUCT datasets using Meshmixer and 3D Slicer is shown in Fig. 1. For this example, the original model and surface texture (totaling 87 MB) was converted to a CT dataset with pixel spacing and slice thickness of 0.5 mm, resulting in 773 slices (totaling 802 MB) over a scan range of 386.5 mm (superior-inferior), and a structure set including the body, treatment volume and mold (totaling 1.8 MB).

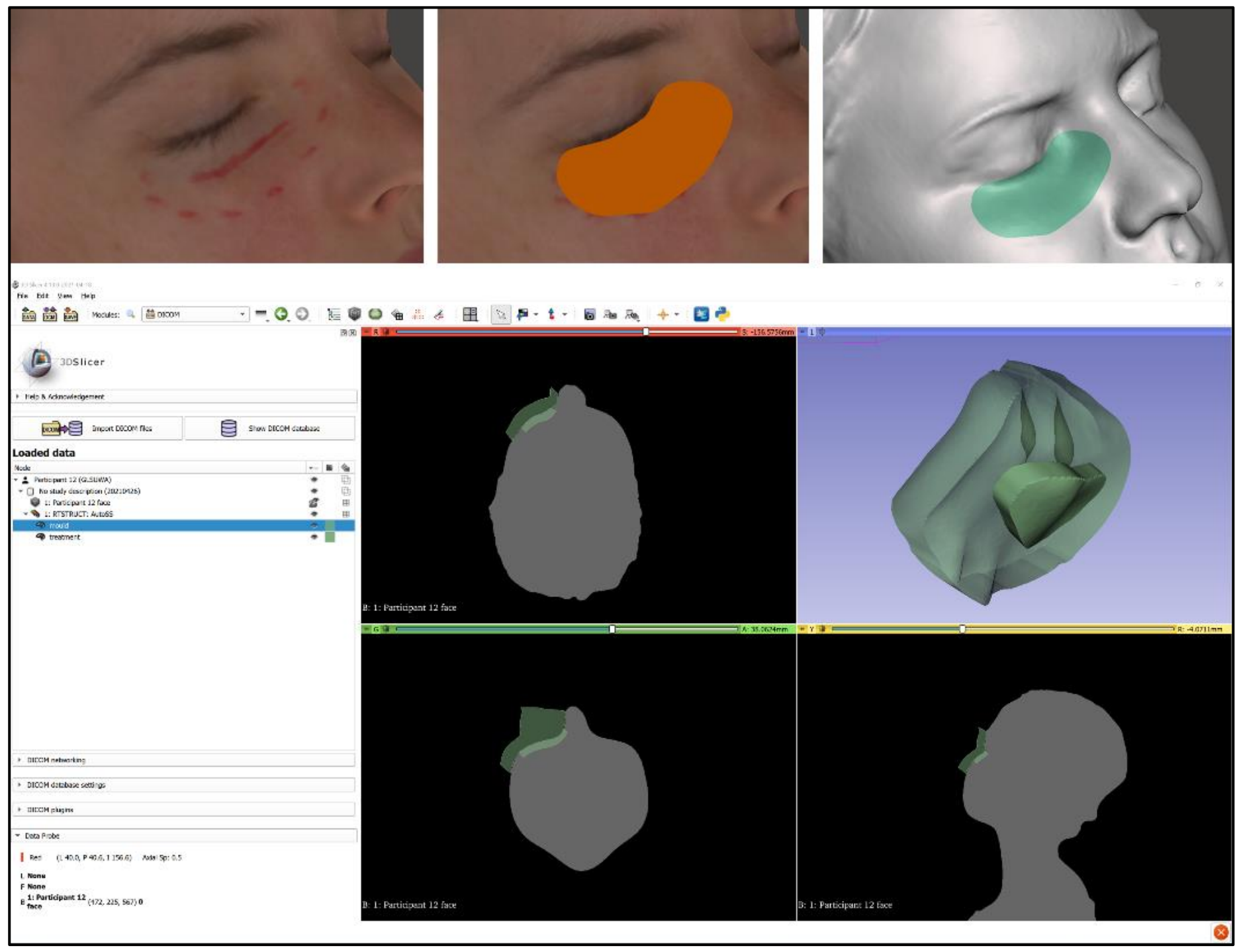

**Fig. 1.** Example production of DICOM CT and DICOM RTSTRUCT data from surface scan

Examples of reconstructed models, mold applicators and planned dose distributions are shown in Fig. 2. Conformal dose coverage of the treatment volume was achieved with the approximately 5 mm separation between multiple catheter channels, and between catheter channels and the skin surface.

An example mold sliced for printing on an Ultimaker 2 Extended+ printer (Ultimaker, Netherlands) using Ultimaker Cura is shown in Fig. 3.

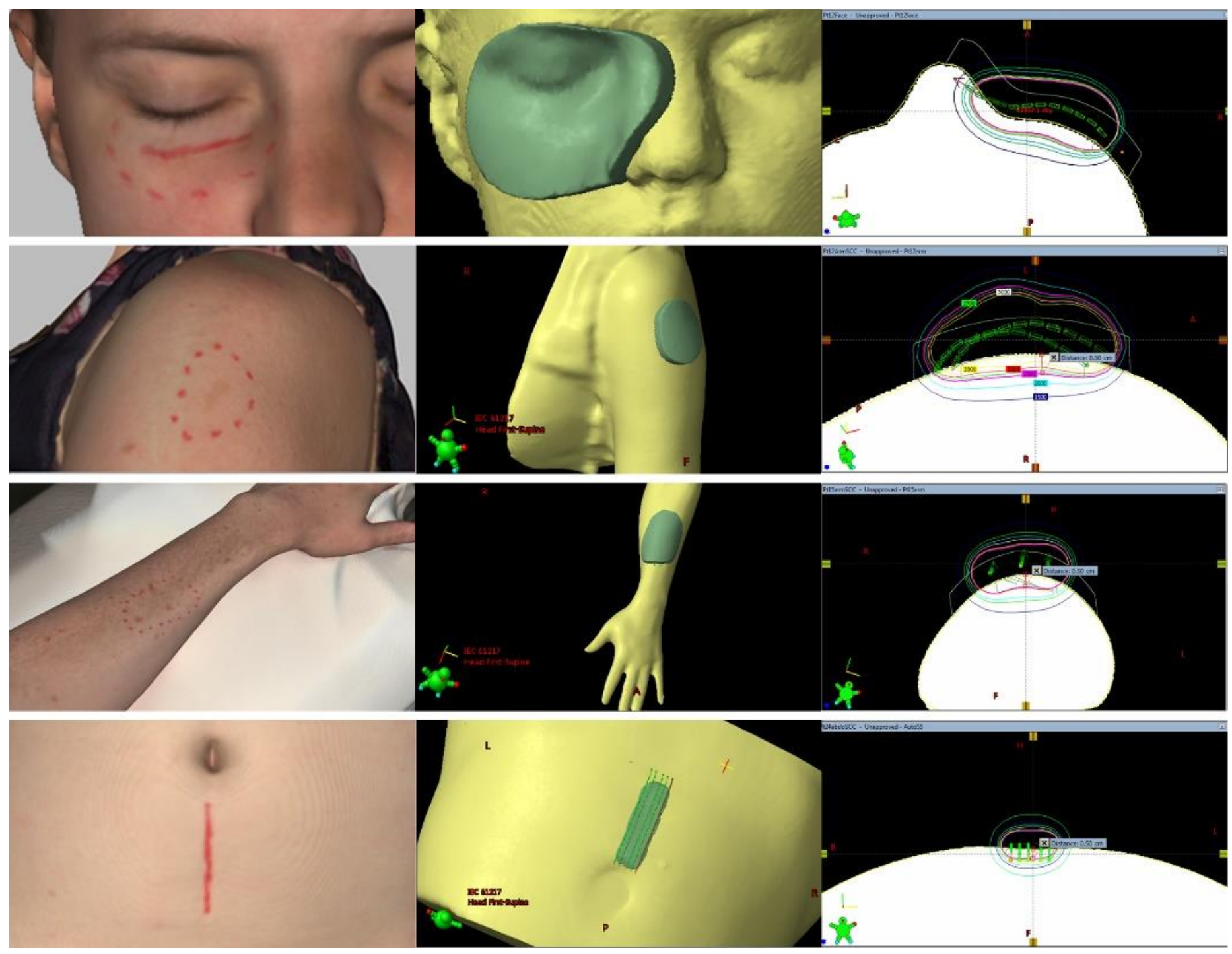

**Fig. 2.** Example brachytherapy surface molds and treatment plans

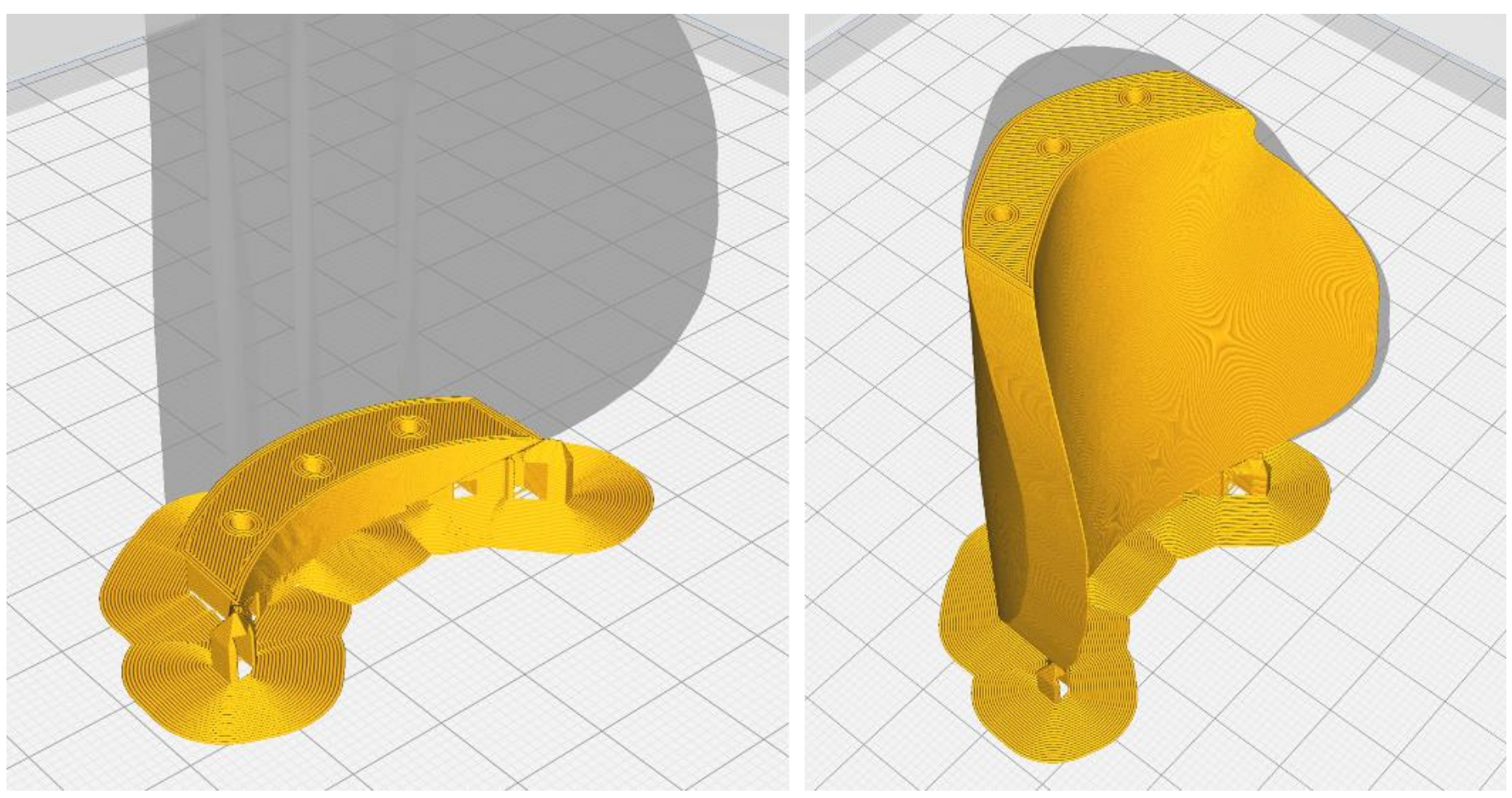

**Fig. 3.** Preview of an example arm mold sliced for 3D printing

## 4 Discussion

Most 3D-printed patient-fitted devices used for radiotherapy treatments are currently designed using medical images, such as simulation CT images. Hard surface modelling tools such as Meshmixer can be used to design simple uniform thickness devices, like external beam radiotherapy bolus, from optically scanned data. However the synthetic water-equivalent CT workflow described in this study allows the optimization of device design within a treatment planning system with consideration of dosimetric impact of device geometry, making it suitable for complex devices such as surface molds.

The lack of internal anatomy did not compromise dose calculations, due to the clinical use of a TG43 based dose calculation algorithm [12]. If 3D printed for clinical use, manufactured molds would be imaged in-situ during CT simulation, allowing for conventional catheter reconstruction and the re-optimization of dwell times for clinical dose calculation and plan approval. CT imaging could also be used for quality assurance of the fit on the patient, geometric dimensions, and radiological properties of the device [16]. Model-based dose calculation algorithms [17] could be used to correct for non-water-equivalence of the 3D printing material or to correct for the lack of full scatter conditions, as required.

The proposed workflow requires an optical scan of the patient, which can be acquired within a few minutes (e.g. at an existing patient consultation prior to CT simulation or treatment), and utilizes free or open source software in addition to any brachytherapy treatment planning system that supports the DICOM standard.

## 5 Conclusion

Optical surface scanning technology can be used for the design of complex 3D-printed patient-specific treatment devices, such as superficial brachytherapy molds, within treatment planning systems by production of simple synthetic CT datasets.

## References

1. Guinot, J. L., Rembielak, A., Perez-Calatayud, J-P., et al.: GEC-ESTRO ACROP recommendations in skin brachytherapy. Radiother. Oncol. 126(3), 377-385 (2018). https://doi.org/10.1016/j.radonc.2018.01.013
2. Stephens, H., Deans, C., Schlect, D., Kairn, T..: Development of a method for treating lower-eyelid carcinomas using superficial high dose rate brachytherapy. Phys. Eng. Sci. Med. 43(4): 1317-1325 (2020). https://doi.org/10.1007/s13246-020-00935-7
3. Song, W. Y., Robar, J. L., Morén, B., Larsson, T., Tedgren., Å, C., Jia, X.: Emerging technologies in brachytherapy. Phys. Med. Biol. 66: 23TR01 (2021). https://doi.org/10.1088/1361-6560/ac344d
4. Bellis, R., Rembielak, A., Barnes, E., Paudel, M., Ravi, A.: Additive manufacturing (3D printing) in superficial brachytherapy. J. Contemp. Brachytherapy 13(4): 468-482 (2021). https://doi.org/10.5114/jcb.2021.108602
5. Arenas, M., Sabater, S., Sintas, A., Arguís, M., et al.: Individualized 3D scanning and printing for non-melanoma skin cancer brachytherapy: a financial study for its integration into

clinical workflow. J. Contemp. Brachytherapy 9(3): 270-276 (2017). https://doi.org/10.5114/jcb.2017.68134

6. Crowe, S. et al.: Evaluation of optical 3D scanning system for radiotherapy use. J. Med. Radiat. Sci., in press (2022). https://doi.org/10.1002/jmrs.562
7. Douglass, M. J. J., Santos, A. M. C.: Application of optical photogrammetry in radiation oncology: HDR surface mold brachytherapy. Brachytherapy 18(5): 689-700 (2019). https://doi.org/10.1016/j.brachy.2019.05.006
8. Bridger, C. A., Douglass, M. J. J., Reich, P. D., Santos, A. M. C.: Evaluation of camera settings for photogrammetric reconstruction of humanoid phantoms for EBRT bolus and HDR surface brachytherapy applications. Phys. Eng. Sci. Med. 44: 457-471 (2021). https://doi.org/10.1007/s13246-021-00994-4
9. Bridger, C. A., Reich, P. D., Santos, A. M. C., Douglass, M. J. J.: A dosimetric comparison of CT- and photogrammetry- generated 3D printed HDR brachytherapy surface applicators. Phys. Eng. Sci. Med., in press (2022). https://doi.org/10.1007/s13246-021-01092-1
10. Schreiber S., Reitemeier, B., Herrmann, T., Fichtner, D., Walter, M.: A Process for Making Cutaneous Radiation Applicators Based on Digital Data, Strahlenther. Onkol. 182: 349-352 (2006). https://doi.org/10.1007/s00066-006-1537-5.
11. Pashazadeh, A., Boese, A., Friebe, M.: Surface anatomy leading to personalized surface applicator: 3D printing for brachytherapy of skin tumors. Transactions on Additive Manufacturing Meets Medicine 1: S03P21 (2019). https://doi.org/10.18416/AMMM.2019.1909S03P21
12. Rivard, M. J., Coursey, B. M., DeWerd, L. A., Hanson, W. F. Huf, M. S., et al.: Update of AAPM Task Group No. 43 Report: A revised AAPM protocol for brachytherapy dose calculations. Med. Phys. 32(3), 633-674 (2004). https://doi.org/10.1118/1.1646040
13. Skinner, L., Knopp, R., Wang, Y-C., Dubrowski, P.: CT-less electron radiotherapy simulation and planning with a consumer 3D camera. J. Appl. Clin. Med. Phys. 22(7), 128-136 (2021) https://doi.org/10.1002/acm2.13283
14. Dipasquale, G., Poirier, A., Sprunger, Y., Uiterwijk, J. W., Miralbell, R.: Improving 3D-printing of megavoltage X-rays radiotherapy bolus with surface-scanner. Radiat. Oncol. 13, 203 (2018). https://doi.org/10.1186/s13014-018-1148-1
15. Mason, D.L., et al.: pydicom: An open source DICOM library (2021). https://doi.org/10.5281/zenodo.4313150
16. Kairn, T., Talkhani, S., Charles, P. H., Chua, B., Yin, C. Y., et al.: Determining tolerance levels for quality assurance of 3D printed bolus for modulated arc radiotherapy of the nose. Phys. Eng. Sci. Med. 44, 1187-1199 (2021). https://doi.org/10.1007/s13246-021-01054-7
17. Beaulieu, L., Tedgren, A. C., Carrier, J-F., Davis, S. D., Mourtada, F., et al.: Report of the Task Group 186 on model-based dose calculation methods in brachytherapy beyond the TG-43 formalism: Current status and recommendations for clinical implementation. Med. Phys. 39(10), 6208-6236 (2012). https://doi.org/10.1118/1.4747264